# Realizing high Near-Room-Temperature Thermoelectric Performance in n-type $Ag_2Se$ through Rashba Effect and Entropy Engineering

Raju K Biswas[1] and Swapan K Pati*[1,2]

[1]Theoretical Sciences Unit
[2]School of Advanced Materials
Jawaharlal Nehru Centre for Advanced Scientific Research (JNCASR)
Jakkur P.O., Bangalore 560064, India. *E-mail: pati@jncasr.ac.in

## ABSTRACT

Although there are enormous numbers of high-temperature thermoelectric materials present, designing a near-room-temperature especially n-type thermoelectric material with high zT is extremely challenging. Generally, pristine $Ag_2Se$ exhibits unusual low thermal conductivity along with high electrical conductivity and Seebeck coefficient, which leads to high thermoelectric performance (n-type) at room temperature. Herein, we report a pseudoternary phase, $Ag_2Se_{0.5}Te_{0.25}S_{0.25}$, which shows improved thermoelectric performance (zT ~ 2.1 at 400 K). Density functional theory reveals that the Rashba type of spin-dependent band spitting originated because of Te-doping, enhancing carrier mobility. Using density functional perturbation theory, we hereby realize that the intrinsic carrier mobility is not only controlled by carrier effective mass, neither deformation potential theory, instead it is substantially limited by longitudinal optical phonon scattering. In fact, locally off-centered S atoms and rising configurational entropy via substitution of Te and S atoms in $Ag_2Se$ significantly reduce the lattice thermal conductivity ($k_{lat}$ ~ 0.34 at 400 K). In order to accurately obtain electrical as well as thermal transport coefficient, we adopt deformation potential theory based on Boltzmann transport formalism. The combined consequence of the Rashba effect coupled with configurational entropy synergistically results in such high thermoelectric performance with the development of new n-type thermoelectric material working at the near-room-temperature regime.

# INTRODUCTION

Recent issues on rising temperature due to global warming emphasize the inhibition of carbon emissions and strengthen energy management and sustainability measures to alleviate the effects of climate change.[1-2] Therefore, research on renewable energy sources such as wind, solar and geothermal energy sources have attracted much attention for the last two decades. Especially, geothermal energy is the internal heat generated in the Earth's interior near-room-temperature regime (150-400 °C), which can be widely used to recover waste heat and utilise it for power generation.[3-4] However, the extensive use of thermoelectric materials is limited by low conversion thermoelectric efficiency. It is related to dimensionless quantity i.e. figure of merit $zT = \frac{S_b^2 \sigma}{k}T$, where $S_b$, σ, κ and T are the Seebeck coefficient, electrical conductivity, thermal conductivity, and absolute temperature, respectively.[5] Interestingly, the efficient way to enhance the zT value is determined by a combination of two major factors; electrical transport should be high, characterised by power factor ($S_b^2\sigma$) and glass-like thermal transport. Some of the compounds which demonstrate high zT at mid-range temperature have been achieved in lead chalcogenide, which not only contains toxic lead (Pb) but also encompasses less abundant tellurium (Te).[6] The scarcity of Te implies to search for new earth-abundant elements that could lead to large zT[7] near-room-temperature through innovative approaches such as defect structure engineering[8], band convergence[9], nano-structuring[10], and phonon glass electron-crystal[11].

Presently, entropy-engineering has become another unique strategy for the reduction of lattice thermal conductivity ($k_{lat}$) of a crystalline solid by manipulating configurational entropies. Configurational entropy increases when multiple components occupy identical atomic sites inside the crystal. Multicomponent chalcogenide like $(SnTe)_{1-2x}(SnSe)_x(SnS)_x$, $(GeTe)_{1-2x}(GeSe)_x(GeS)_x$ and $(PbTe)_{1-2x}(PbSe)_x(PbS)_x$ are very well known pseudoternary systems to exhibit low lattice thermal conductivity. It occurs because an increase in configurational entropy arises from point defect phonon scattering, mass fluctuation of different element and strain-induced resulting from atomic size mismatch.**[12-16]** Additionally, the large atomic size mismatch between the constituent atoms could also lead to locally strained in the global structure. Moreover, the large size mismatch of Te and S atoms would generate off-centering of S atoms which introduces local dipole moment and soft phonon modes in the crystal. As a result, the soft phonons effectively scatter heat-carrying acoustics phonons and reduces lattice thermal conductivity significantly.**[17]** Along with entropy engineering, the Rashba effect has an additional effect in the fields of thermoelectricity through the spin-orbit coupling (SOC) mechanism in the charge carrier transport phenomenon.**[18]** The Rashba spin splitting leads to a distinctive feature in the electronic density of states (DOS) due to the unique Fermi surface topology **[19]**. Consequently, the power factor ($S_b^2\sigma$) is found to be large because of the sharp peak in DOS **[20]**. Thus, the Rashba spin splitting can potentially be a very effective mechanism for thermoelectricity.

Chalcogenide families such as PbTe and its solid solution with PbSe and PbS are well-known semiconductors to show excellent thermoelectric efficiency.**[21]** However, PbTe based materials are not eco-friendly because of the presence of Pb, and as a result, the use of PbTe remains unsuitable for mass-market applications. Various other thermoelectric materials such as SiGe,**[22]** skutterudites,**[23]** half-Heusler compounds,**[24]** colusites,**[25]** SnSe,**[26]** SnS,**[27]** $Cu_2Se$,**[28]** exhibit high zT (1.5-2.5) values exclusively above 500 K**.** Interestingly, none of the above materials show zT above 1.0 at room temperature except the only compound $Bi_2Te_3$ [zT~1.3-1.7]. Although the performance of $Bi_2Te_3$ has been strategically improved via various mechanisms, it is indeed coming out to be p-type.**[29]** In Contrast, the n-type $Bi_2Te_3$ remains below zT~1.0 at room temperature.**[30]** Meanwhile, it has become imperative to explore new n-type thermoelectric materials that exhibit zT above unity near-room-temperature as maximum energy wastage in line with geothermal power generation occurs approximately in this temperature range.**[31]**

The current materials with high TE performance working in near-room-temperature range mainly include $Bi_2Te_3$-based systems, $Mg_3Sb_2$, MgAgSb, $Mg_2$(Si, Sn), $AgSbTe_2$ and $\beta$-$Ag_2Q$ (Q=Se, Te)**[32-33].** Recently, there has been numerous research for a long time where $\beta$-$Ag_2Q$ is considered as a promising candidate for thermoelectric application because of its intrinsic metal-like electrical conductivity and inherently glass-like thermal conductivity at room temperature.[32,34] Experimentally, Silver selenide specimens were prepared by a direct reaction of the source elements (6N purity) in evacuated (≈$10^{-4}$ Torr) quartz tubes and X-ray diffraction analysis confirmed the formation of β - $Ag_2Se$.[34] However, this material shows discrepancies in zT at the temperature range 300-400K, which have been reported in the literature.[35-36] Subsequent efforts were attempted towards increasing the figure of merits of thermoelectric materials.[37] Finally, the achievement of a high zT > 1.0 at temperature range (300-375 K) is demonstrated by improving charge carrier mobility along with tuning the carrier concentration via the addition of minute anion excess.[38] In due course, the synthesis of solid solution $Ag_2Se_{1-x}Te_x$ has been confirmed by X-ray diffraction data and Rietveld refinement.[39] In fact, Drymiotis *et. al.* showed $Ag_2Se-As_2Te$ phase diagram at different compositions using density functional theory (DFT). In the present study, we have further doped the Sulphur atom in the ternary phase and examine the thermoelectric performance of the pseudoternary phase, $Ag_2Se-Ag_2Te-Ag_2S$. Our analysis encompasses the comprehensive understanding in calculating intrinsic carrier mobility, not only limited by longitudinal acoustics phonon scattering, rather governed by longitudinal optical phonons. Achieving the excellent thermoelectric performance near-room-temperature range in an $Ag_2Se$-rich phase, $Ag_2Se_{1-2x}Te_xS_x$ by minimizing lattice thermal conductivity ($k_{lat}$) using the mechanism such as configurational entropy engineering is demonstrated. Furthermore, we will also discuss an increment of power factor due to the Rashba type of spin splitting and as a result, we achieve a very high value of zT~2.1 at 400 K in $Ag_2Se_{0.5}Te_{0.25}S_{0.25}$.

**THEORETICAL METHODS**

**(a) First-principles simulation:**

First-principles based density functional theory calculations have been performed by a Quantum Espresso package[40] which employs the pseudopotential model considering the ion-electron interaction inside the material. We use full relativistic projected augmented wave(PAW) pseudopotentials[41-42] as exchange-correlation energy of the electrons with generalized gradient approximated (GGA)[43] functional parametrized by Perdew, Burke, Erzenhoff (PBE).[44] The crystal structure is fully relaxed using the conjugate gradient scheme until the magnitude of Helmann-Fynman force becomes less than 0.025 eV/Å. The electronic wave function and charge density cutoff are taken to be 70 Ry and 400 Ry, respectively. Brillouin zone integration of orthorhombic crystal structure is sampled on the uniform grid of 24 x 16 x 16 k-points. A strict energy convergence of $10^{-8}$ eV has been used to obtain phonon frequencies accurately. The sharp discontinuity of the electronic states near the Fermi level is smeared out with the Fermi-Dirac distribution function with a broadening of 0.003 Ry.

We consider the orthorhombic phase of $Ag_2Se$, $\beta$-$Ag_2Se$ that undergoes a first-order phase transition to a cubic phase around 407 K. The low-temperature phase is a narrow bandgap semiconductor with space group $P2_12_12_1$ and lattice parameters a=4.35 Å, b=7.07 Å and c=7.78 Å.[45,46] The unit cell of $\beta$-$Ag_2Se$ consist of 12 atoms with four molecules of $Ag_2Se$ having 8 Ag and 4 Se atoms. Under high temperature above 407 K, it transforms to a cubic phase where materials become metallic with Ag atoms behaving as superionic characteristics.[47] We also incorporate the spin-orbit coupling (SOC) effect while calculating band structure. It has been observed that the effect of SOC is very prominent for doped $Ag_2Se$. Pristine, S-doped, Te-doped and co-doped (Te & S) $Ag_2Se$ have been simulated to obtain electronic band structure along the high symmetry points. Furthermore, phonon dispersion properties of pristine and doped $Ag_2Se$ have been calculated using density functional perturbation theory (DFPT). $Ag_2Se$ unit cell containing 12 atoms, is considered for phonon frequency calculation. A uniform grid of 2 x 2 x 2 q-mess is taken into consideration. To deeply understand the origin of ultralow lattice thermal conductivity, we further focused on various branches in the phonon dispersion curve of both pristine and doped $Ag_2Se$. In addition to that, we have also computed configurational entropy or mixing of entropy using the formula:

$$\Delta S_{mix} = -nR(x_1 \ln x_1 + x_2 \ln x_2) \quad (1)$$

where $x_1$ and $x_2$ are mole fraction of species 1 and 2 respectively.

**(b) Carrier Transport Properties Study using Boltzmann Formalism:**

The Seebeck Coefficient is defined as[48,49]:

$$S_b = -\frac{\pi^2}{3e} k_B^2 T \frac{\partial ln(\sigma(E))}{\partial E} \quad (2)$$

here, $k_B$: Boltzmann constant; e: electronic charge, T: absolute temperature and $\sigma(E)$: energy-dependent density of states (DOS). On the other hand, the electronic transport properties are mostly governed by carrier mobility which is strongly influenced by charge carrier effective mass. The well-known mobility expression based on deformation potential theory (DPT) proposed by Bardeen and Shockley in 1950, where the longitudinal acoustic (LA) phonon limited mobility of a three-dimensional (3D) semiconductor was given as[50]:

$$\mu_{LA} = \frac{2^{3/2}\pi^{1/2}}{3} \frac{C_{3D}\hbar^4 e}{(E_\beta^c)^2 m^{*5/2}_{(e/h)} (k_B T)^{3/2}} \quad (3)$$

where $m^*_{(e/h)}$ is the effective mass of charge carriers (electrons and holes), $C_{3D}$ is the elastic constant, and $E_\beta^c$ is the deformation potential constant that is defined as the energy shift of the band edge position with respect to the uniaxial strain. We can also estimate electronic relaxation using the expression $\mu_{LA} = \frac{e\tau_{LA}}{m^*}$ The elastic constant is derived as:

$$2(E - E_0) = C_{3D} S_0 (\Delta l/l_0)^2 \quad (4)$$

where $E_0$ and $S_0$ are the energy and lattice volume of the unit cell without strain. This parabolic equation is employed to compute the elastic constant, $C_{3D}$. Deformation potential constant can also be calculated using the expression:

$$E_{\beta}^{c} = \frac{\Delta E_{band}}{(\Delta l/l_0)} \qquad (5)$$

where $\Delta E_{band}$ is a shift in energy level at the band edge position because of lattice strain $\Delta l/l_0$.

To calculate mode resolved intrinsic carrier mobility limited by longitudinal optical (LO) phonon, we recall small-momentum behaviour of the Fröhlich interaction is well described by the leading order in Vogl's model for 3D semiconductor[51]:

$$g_{Fr}^{3D}(q_p) = \frac{4\pi e^2}{V|q_p|\epsilon_p^q} \sum_a \frac{e_{q_p} \cdot Z_a^b \cdot e_{q_p LO}^a}{\sqrt{2M_a \omega_{q_p LO}}} \qquad (6)$$

where $e$ is the electronic charge, V is the unit cell's volume, $\epsilon_p^q$ is the bulk dielectric constant, and $e_{q_p} = \frac{q_p}{|q_p|}$. $M_a$ is the atomic mass, $Z_a^b$ is Born effective charge associated to particular atom a, $\omega_{q_p LO}$ is the longitudinal optical phonon frequency at wave vector $q_p$. Thus the total scattering rates due to LO phonon can be written in terms of Fröhlich interaction as follows[52]:

$$\frac{1}{\tau_{LO}} = \frac{V}{4\pi^2} \frac{2\pi}{\hbar} \int_0^{2\pi} d\phi \int_{q_{min}}^{q_{max}} |g_{Fr}^{3D}(q_p)|^2 \frac{1}{\hbar\omega_{q_p}} q_p (n_{\hbar\omega_{q_p}} + \frac{1}{2} \mp \frac{1}{2}) \, dq \qquad (7)$$

Once we determine $\tau_{LO}$, we can evaluate longitudinal optical phonon limited carrier mobility using the expression $\mu_{LO} = \frac{e\tau_{LO}}{m^*}$. Using Matthiessen's rule $\frac{1}{\mu} = \frac{1}{\mu_{LA}} + \frac{1}{\mu_{LO}}$, we obtain effective carrier mobility ($\mu$). Finally, the electrical conductivity can be calculated using the Drude formula $\sigma = ne\mu$, where n: carrier concentration, e: electronic charge.

The important parameter to calculate lattice thermal conductivity is phonon group velocity, $\vartheta_{iq}$ ($\vartheta_{iq} = \nabla(E_{iq})/h$), which are obtained from first-principles calculations. The phonon relaxation time has been computed by solving Boltzman transport equations within relaxation time approximation (RTA). We consider the phonon relaxation time, $\tau_{iq}$, using the collision term in Boltzmann Transport Equations (BTE)[53]:

$$\frac{1}{\tau_{iq}} = k_B T \frac{4\pi^2 (E_{ph})^2}{hC_{3D}} \sum_{j,q' \in BZ} \left\{ (1 - \frac{v_{jq'}}{v_{iq}})\delta[E_{iq} - E_{jq'}] \right\} \quad (8)$$

Where $E_{ph}$ is deformation potential constant for phonon (phonon-phonon (acoustics) scattering strength) which is defined as the change in energy of the longitudinal acoustic phonons in the linear region near high symmetry Γpoint per unit lattice strain ($\Delta l/l_0$). The computed relaxation time $\tau_{iq}$ is employed to calculate total lattice thermal conductivity ($k$). Finally, $k$ can be expressed as[54]:

$$k = \sum_{i,i',q,q'} C_{iq} \vartheta_{iq} \vartheta_{i'q'} \tau_{iq} \quad (9)$$

where $C_{iq}$ is the specific heat, $\vartheta_{iq}$ is the phonon group velocity and $\tau_{iq}$ is the phonon relaxation time. These parameters have been calculated based on the Boltzmann transport formalism.

**RESULT AND DISCUSSION**

In this work, we have considered the orthorhombic phase of Ag$_2$Se (β-Ag$_2$Se) whose full description of its crystal structure was studied by Wiegers in 1971 based on X-ray powder diffraction data[55]. We notice that all of the crystallographic sites corresponding to Ag atoms are not equivalent. This particular phase contains crystallographically two distinct silver atoms, Ag1 is coordinated tetrahedrally, whereas Ag2 atoms are placed in almost triangular symmetry[**Shown in Fig. S1**, supporting information]. Generally, Ag$_2$Se seems to be a non-stoichiometric compound at both low as well as high-temperature phases. In fact, the low-temperature phase of Ag$_2$Se reveals not only excellent thermoelectric efficiency even at room temperature but rather is also an n-type semiconductor. Despite numerous research on thermoelectric properties of Ag$_2$Q (Q=Te, Se) reported previously,[32-35] to date, there is no significant studies on pseudoternary (Ag$_2$Se-Ag$_2$Te-Ag$_2$S) system for thermoelectric application because of its complex chemistry. Here, we discuss the structural, thermodynamical, chemical aspect and thermoelectric properties of the pseudoternary phases of Ag$_2$Se0.

Recently, *ab-initio* based density functional perturbation theory (DFPT) methods have become well-established techniques to investigate the structural stability of various materials.[56,57] In this regard, our lattice dynamics study using phonon dispersion calculation reveals that pristine and doped Ag$_2$Se are dynamically stable. In phonon dispersion curves shown in **Fig. S2**, supporting information, we observe all vibrational modes are positive in magnitude which implies mechanical stability of both pristine and doped Ag$_2$Se crystal structures. Furthermore, thermodynamic stability is also examined performing formation energy and binding energy calculations for doped Ag$_2$Se. The estimated formation energies for S-doped, Te-doped and Te & S co-doped Ag$_2$Se are found to be -0.81 eV, -0.27 eV and -0.40 eV, respectively (Tabulated in **Table S1**, supporting information). Moreover, binding energies are also calculated which turns out to be -6.12 eV, -5.02 eV and -10.47 eV, respectively (Tabulated in **Table S1**, supporting information). Interestingly, formation energy in combination with binding energy determines the chemical stability of these pseudobinary and pseudo-ternary phases of Ag$_2$Se.

In order to determine electronic structure properties, we use density functional theory (DFT) of Te and S co-doped Ag$_2$Se, Te-doped Ag$_2$Se, S-doped

Ag$_2$Se, and undoped Ag$_2$Se. **Fig. 1** represents electronic band dispersion of pristine, S-doped, Te-doped and Te & S co-doped, Ag$_2$Se, respectively along the high symmetry points (See **Table S2**, supporting information). The bandgap estimated in PBE level for pristine is well consistent with other theoretical calculations.**[55]** It is interesting to notice Sulphur dopping widen the bandgap which is also shown in **Fig. 1(b)**. In contrast, Te-doping narrows down the band gap between valence band maxima (VBM) and conduction band minima (CBM) and most importantly lifts the spin degeneracies due to spin-orbit coupling (SOC) originated because of the presence of heavy nuclei in the unit cell (See **Fig. 1(c)**). Finally, **Fig. 1(d)** indicates that Te & S co-doping serves the combined purpose simultaneously that not only opens up the band gap between VBM and CBM but also lifts spin degeneracy near the Fermi level which is effective to enhance electrical conductivity. In fact, we compare bandgap for pristine and doped Ag$_2$Se and show in **Fig. S3**, supporting information. Moreover, we also calculate the HSE06 bandgap and tabulate it in **Table S3**, supporting information.

Fundamentally, the SOC driven spin-dependent band structure spitting is known as the Rashba effect. In fact, the indirect bandgap along with more number of energy bands near the Fermi level originated due to the Rashba effect would eventually be beneficial for thermoelectric performance. To understand the Rashba effect rigorously, we have calculated band structures for pristine Ag$_2$Se and doped Ag$_2$Se as shown in **Fig. 2**. The k-dependent spin-splitting of the conduction band shifts the band edge from the high symmetry Γ-point due to Tellurium doping along both K → Γ and M → Γ directions. The momentum offset, Δk, is the difference between the Γ-point and shifted band-extrema in k-space. For Te-doped Ag$_2$Se and Te & S co-doped Ag$_2$Se, Δk in the conduction band along M → Γ high symmetry directions are 0.025 and 0.037 Å$^{-1}$, respectively. Whereas, Δk for such splitting in its valence band is found to be the same as 0.025 Å$^{-1}$ along the M → Γ high symmetry directions for single doping and co-doping cases (see **Figs. 2(c) and 2(d)**). Therefore, we conclude from the above discussion that the momentum offset differing extents of Δk in the frontier bands create an indirect bandgap for co-doped (Te & S) Ag$_2$Se leads lowering in electron-hole recombination rates(see **Table 1)**. However, the extent of splitting in the conduction band for S-doped Ag$_2$Se along the K → Γ and M → Γ directions is significantly very small compared to that Te doped (see **Table 1**). Interestingly, the band-splitting observed for

S-doped $Ag_2Se$ could be not because of SOC, rather it is due to off-centering of S atoms from its equilibrium position(shown in **Fig. 3**). Through off-centering the S atom from its original position, the crystal further reduces energy by 2.09 meV. The strength of the Rashba effect can be calculated using the parameter $\alpha = E_R/2\Delta k$, where $E_R$ is the amplitude of the band splitting at the band edge in a particular direction. The estimated α value for co-doped $Ag_2Se$ in the M → Γ direction 1.004 eV.Å which is of the same order as those reported by previous computational studies on $Ag_2Te$.**[55]**

To further examine the consequence of the Rashba effect on electronic structure, we plot the density of states (DOS) in **Fig. 4(a)**. Surprisingly, we find that the total DOS shows a sharp peak near the Fermi level because of Te doping in the unit cell. In fact, pristine and S-doped $Ag_2Se$ does not exhibit such an extent of the peak near the Fermi level. Consequently, a sudden peak in the total DOS as a result of tellurium doping gives rise to DOS effective mass, contributing to enhancement in the Seebeck coefficient. We have also studied atomic orbital projected DOS (pDOS) to understand the orbital contribution due to the Rashba effect. In fact, it has been confirmed from the pDOS in **Fig. 4(b)** that tellurium ions have the highest orbital contribution near the Fermi level and consequently Te's are mainly responsible for Rashba types of effect.

Finally, we perform transport calculations based on favourable electronic band structure to evaluate Seebeck coefficient ($S_b$), electrical conductivity (σ) and thermoelectric power factor ($S_b^2\sigma$) for pristine and doped $Ag_2Se$. As described in the computational section, we solve the Boltzmann transport equation (BTE) within relaxation time approximation (RTA) to estimate the Seebeck coefficient ($S_b$) and mobility(μ). We compute $S_b$ using **equation 2** and plotted in **Fig. 5(a)** within the temperature range (100-400 K). Interestingly, the estimated Seebeck coefficient turns out to be negative which implies n-types carriers are the determining factor for transport in $Ag_2Se$. We now compare the calculated $S_b$ for pristine $Ag_2Se$ (197 μ$V/K$ at 200 K) with the experimentally measured Seebeck coefficient (~175 μ$V/K$ at 200 K).**[34]** Accordingly, the $S_b$ for Te-doped $Ag_2Se$ is calculated to be 155 μ$V/K$ at 400 K which is slightly lower than pristine $Ag_2Se$.

Therefore, we substitute Sulphur in place of Se and the estimated Seebeck coefficient turns out to be 525 $\mu V/K$, 292 $\mu V/K$ at 400 K for S-doped and Te & S co-doped $Ag_2Se$, respectively. Importantly, The $S_b$ (~292 $\mu V/K$ at 400 K) estimated in the case of co-doping is far more than $Ag_2Te$ (~150 $\mu V/K$ at 400 K).[58]

We calculate intrinsic charge carrier mobility considering the fact that carriers are not only scattered by longitudinal acoustic phonons but also limited by low energy longitudinal optical phonons. According to deformation potential theory (DPT), the charge carrier mobility($\mu_{LA}$) largely depends on the carrier scattering information in combination with carrier effective mass and deformation potential constant (see **Table S4, S5**, supporting information). In fact, the carrier effective mass (m*) has been derived from the 2nd order derivative of electronic band dispersion curves with momentum and tabulated in **Table S4, S5** in supporting information. We further compare m* for pristine $Ag_2Se$ with other ab-initio results.[59] Using these parameters, we find temperature-dependent carrier mobility ($\mu_{LA}$) using **equation (3)**. To incorporate longitudinal optical (LO) phonon limited carrier transport, we use the Fröhlich interaction term which is described in **equation (6)**. The carrier-optical phonon scattering strength ($g_{Fr}^{3D}(q_p)$) depends upon Born effective charge ($Z_a^b$), bulk dielectric constant ($\epsilon_p^q$) and magnitude of an optical phonon ($\omega_{q_pLO}$). These parameters can easily be obtained using DFPT implemented in Quantum Espresso distribution. Then the scattering rate can be computed using **equation (7)** and finally calculate optical phonon limited intrinsic mobility ($\mu_{OP}$). Finally, the estimated μ is coming out to be of the order of $10^3$ for doped $Ag_2Se$ at temperature 400 K. The mobility for pristine $Ag_2Se$ evaluated considering the scattering model driven by both acoustic as well as optical phonons, is eventually well consistent with experimental results.[34] In fact, mobility for Te-doped $Ag_2Se$ is found to be 3082 $cm^2/Vs$ which matches well consistent with the experiment.[39] Once carrier mobility is known (shown in **Fig. 5(b)**), we can calculate carrier electrical conductivity using the Drude formula $\sigma = ne\mu$ and drawn in **Fig. 5(c)**). Here, we find that carrier concentration is of the order of $10^{19}$ $cm^{-3}$. The estimated σ for pristine, S-doped, Te-doped, co-doped

(Te & S) Ag$_2$Se are found to be 2457 S/cm, 432 S/cm, 7499 S/cm, 3160 S/cm, respectively, at room temperature. To validate DPT theory, we compare theoretically predicted σ of pristine Ag$_2$Se (~2457 S/cm at 300 K) with experiment (~1988 S/cm at 300 K).**[34]** The enhancement of electrical conductivity in the case of Te-doping occurs due to the appearance of multiple levels near the Fermi level as a consequence of the Rashba type spin splitting. Since carrier electrical conductivity increases due to Te doping, it has a direct impact on the thermoelectric power factor (PF). As a result, we achieve PF to be 152 $\mu W/cm.K^2$ for Te & S co-doped Ag$_2$Se at 300 K. The thermoelectric power factor for pristine and doped Ag$_2$Se are shown in **Fig. 5(d)**.

**Fig. 6(a)** displays a magnitude of exceptionally lower soft vibrational modes at high symmetry M and K-points in the Brillouin zone. The phonon frequency for pristine Ag$_2$Se at M-point (K-point) is observed to be 7.60 cm$^{-1}$ (13.65 cm$^{-1}$). For S-doped, Te-doped and Te & S co-doped Ag$_2$Se, phonon modes get further softer which eventually effective for good thermoelectric performance. The observed phonon modes (cm$^{-1}$) at M-point (K-point) are found to be 6.82 (12.07), 4.40 (11.7), 4.31 (11.58) for S-doped, Te-doped and Te & S co-doped Ag$_2$Se, respectively. Interestingly, the frequency of the soft phonon mode at M-point for the co-doped system is in the same order of n-type BiSe.**[60]** To unravel the origin of such soft mode in the phonon spectra, we present phonon density of states (DOS), determined from first principles. Total phonon density of states (**Fig. 6(b)**) of Te-doped and Te & S co-doped Ag$_2$Se clearly confirms the sharp peak arising because of Te doping in the low-energy phonon modes which is highlighted by the black ellipse. These low energy acoustics modes are the signature to achieve low lattice thermal conductivity in such a pseudo-ternary phase.

Additionally, entropy engineering is also utilized to modulate lattice thermal conductivity via Te and S doping at Se sites in Ag$_2$Se crystal. As calculated using **equation (1)** and shown in **Table 2**, the configurational entropy ($\Delta S_{conf}$) gradually increases with S-doping, Te-doping and Te & S co-doping, respectively. The calculated $\Delta S_{conf}$ are found to be 5.276 Joule/K-mol, 6.658 Joule/K-mol, 6.976 Joule/K-mol for S-doped, Te-doped and co-doped Ag$_2$Se, respectively. In fact, the order of the $\Delta S_{conf}$ for doped cases is well consistent with another similar study, Te-doped in $Cu_7PSe_6$.**[61]** Eventually, the substitution of Te and S in $Ag_2Se$ increases the configurational entropy and point defects scattering which results in lowing lattice thermal conductivity significantly.

Both pristine and doped $Ag_2Se$ have intrinsically low lattice thermal conductivity ($k_{lat}$) at room temperature due to an increase in configurational entropy dominated by point defect scattering.**[62-63]** Herein, we solve the Boltzmann transport equation (**equation 8**) to evaluate phonon relaxation time at temperature range (100-400 K). The parameters required to calculate phonon relaxation time, have been obtained using first principles (See **Table S6**, supporting information). Pristine $Ag_2Se$ seems to show $k_{lat} \approx 0.83\ W/m.K$ at 400 K which is well consistent with experiments.**[34]** Via Te and S doping at Se site in $Ag_2Se$ crystal, $k_{lat}$ gradually decreases and the $k_{lat}$ is calculated using the **equation 9** for S-doped, Te-doped and co-doped (Te & S) $Ag_2Se$ which are found to be 0.8 $W/m.K$, 0.63 $W/m.K$, 0.34 $W/m.K$, respectively at 400 K. The temperature-dependent $k_{lat}$ have been shown in **Fig. 7(a)**.

Finally, the dimensionless parameter, zT as a function of temperature have been computed and shown in **Fig. 7(b)**. The pristine $Ag_2Se$ shows the highest zT value (~ 1.05) at 400 K which is also compared and well consistent with other studies.**[34,64]** The zT is calculated extending temperature up to 400 K because $Ag_2Se$ crystal undergoes structural phase transition near 410 K.**[64]** In fact, we also compare estimated zT for Te-doped $Ag_2Se$ which is found to be 0.74 at 400 K. Interestingly, due to sulphur doping, we achieve maximum zT around 2.1 at 400 K in co-doped (Te & S) $Ag_2Se$.

**Conclusion**

In conclusion, we investigate the superlattice of $(Ag_2Se)_{1-2x}(Ag_2Te)_x(Ag_2S)_x$ where we have thoroughly analyzed electronic as well as thermal transport behaviour of pseudo-ternary phase near-room-temperature. This superlattice shows higher configurational entropy arising due to a large atomic size mismatch between Te & S atoms in Ag$_2$Se unit cell, which would lead to point defect scattering. Due to such defect scattering, we observe a very low value of lattice thermal conductivity (0.34 $W/m.K$ at 400 K). This is because of the fact that phonon scattering eventually occurs due to effective point defect scattering involving a broad range of mass fluctuation of constituting atoms (Se/Te/S). Additionally, the Rashba effect, spin-splitted band dispersion appeared due to Te, introduces a sharp peak in the density of states near the conduction band region just above the Fermi level. Indeed, we find a moderately high thermoelectric power factor in the case of the co-doped superlattice. As a result of such relatively high PF and ultra-low lattice thermal conductivity, we achieve an exceptionally high zT (2.1 at 400 K) in Ag2Se$_{0.5}$Te$_{0.25}$S$_{0.25}$. In fact, the SOC driven spin splitting Rashba effect helps to greatly enhance the thermoelectric power factor and simultaneously enhancement of configurational entropy due to point defect scattering leads to reducing lattice thermal conductivity. Eventually, these two combined effects synergically increase the thermoelectric zT value in the pseudo-ternary phase operating in a near-room-temperature regime.

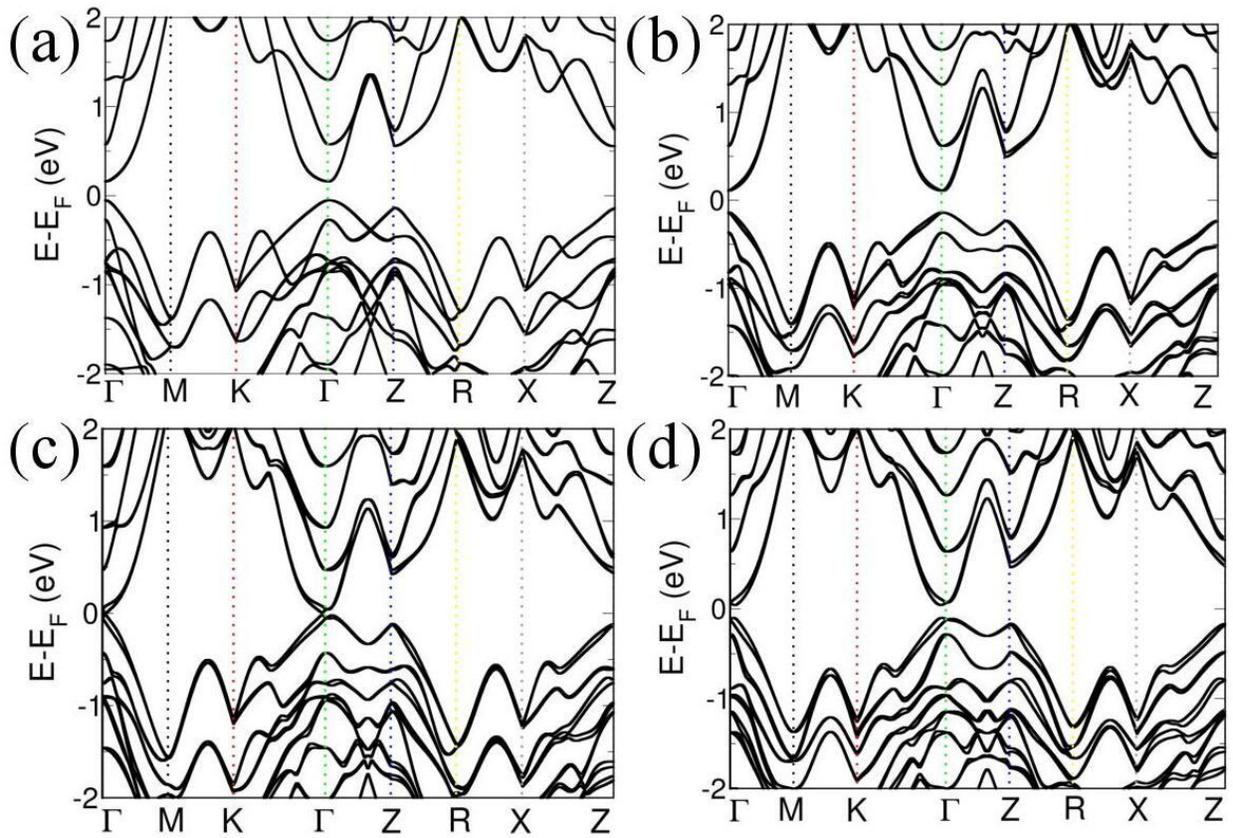

**Fig. 1:** Electronic band structure of (a) Pristine, (b) S doped, (c) Te doped, (d) Te & S co-doped, $Ag_2Se$, along the high symmetry k-points. All the band dispersions are plotted including spin-orbit coupling (SOC) due to the presence of heavy metals in the unit cell. The Fermi level is considered as a reference for each figure.

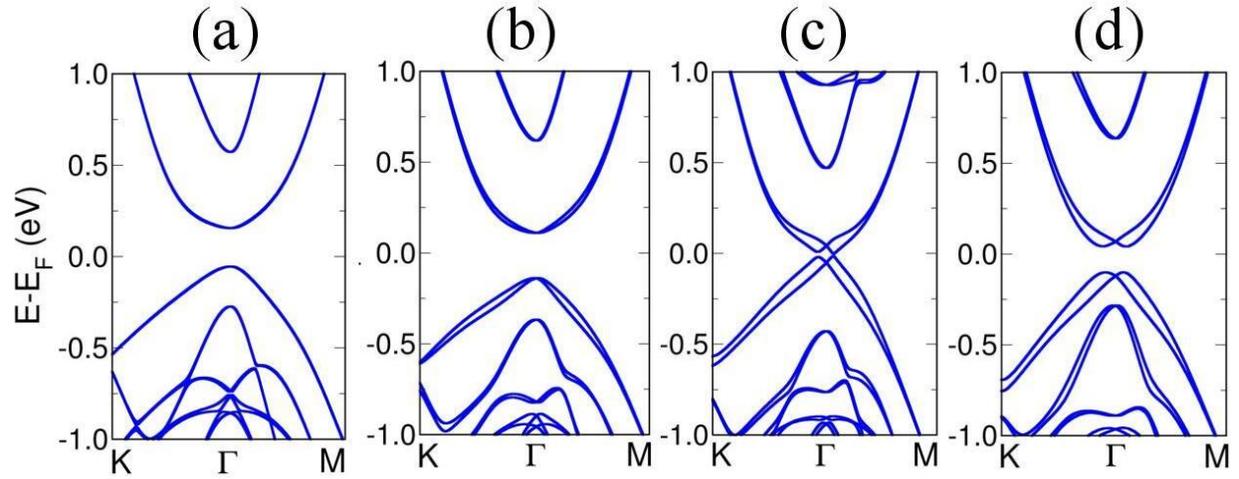

**Fig. 2:** Rashba type effect in $Ag_8Se_4$. The PBE-SOC electronic band structure of (a) $Ag_8Se_4$, (b) $Ag_8Se_3S$, (c) $Ag_8Se_3Te$, (d) $Ag_8Se_2TeS$ have been plotted for a dense k-grid. The Rashba splitting for $Ag_8Se_2TeS$ system in both VBM and CBM along the M → Γ and K → Γ directions are prominent. Interestingly, momentum offset ($\Delta k$) for co-doped systems are not equal for VBM and CBM which transform direct to indirect bandgap and effectively help to reduce electron-hole recombination rates.

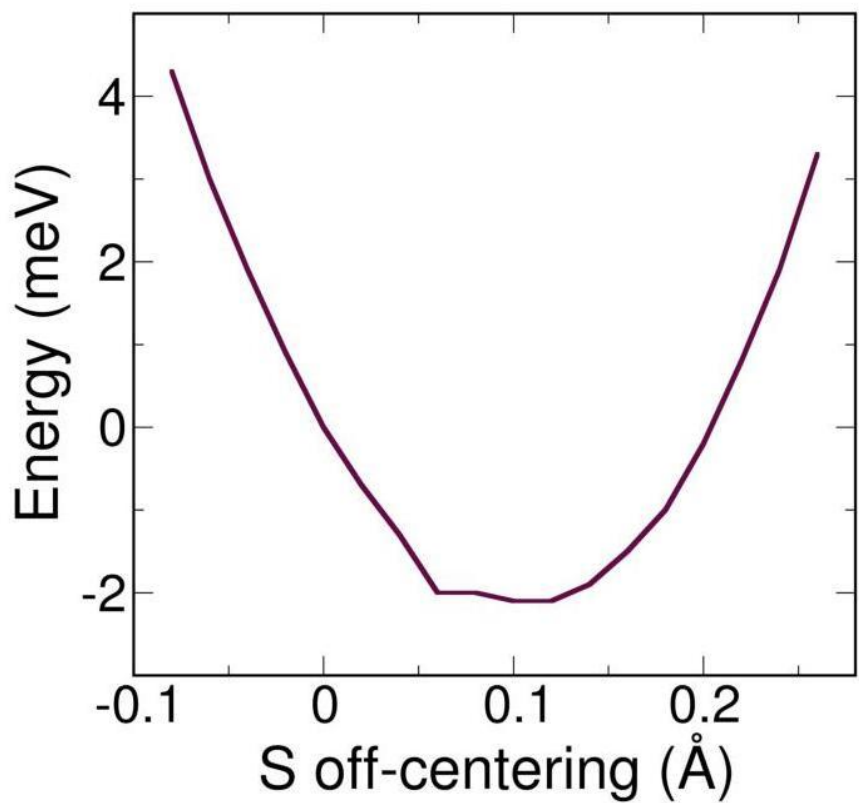

**Fig. 3.** Energy profile vs. coordinate of displacement of sulphur (S) atom from its equilibrium position. Through the displacement (around 0.1 Å) of the S atom, the crystal reduces energy by 2.09 meV.

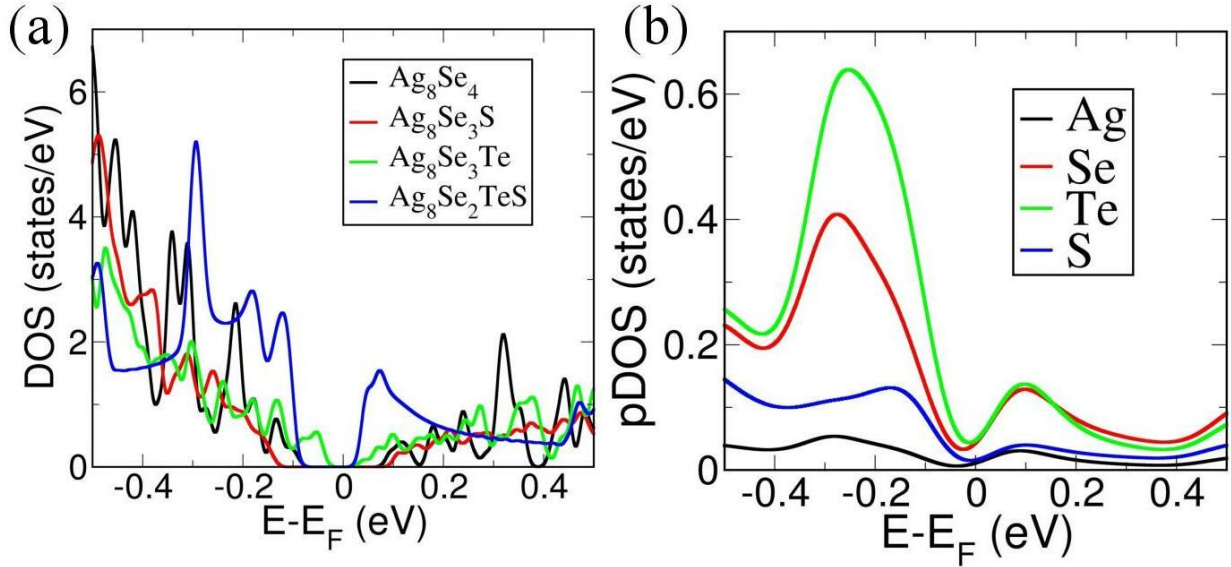

**Fig. 4:** (a) Total density of states (DOS), (b) atom projected DOS (pDOS), have been plotted as a function of energy. Interestingly, we observe sharp peaks just near the Fermi level in DOS for the $Ag_8Se_2TeS$ system (blue dotted line). From pDOS, it is clear that the atom project orbital arising near the Fermi level predominantly originated from the Te atom (represented by green colour). The Fermi level is considered as a reference for each figure.

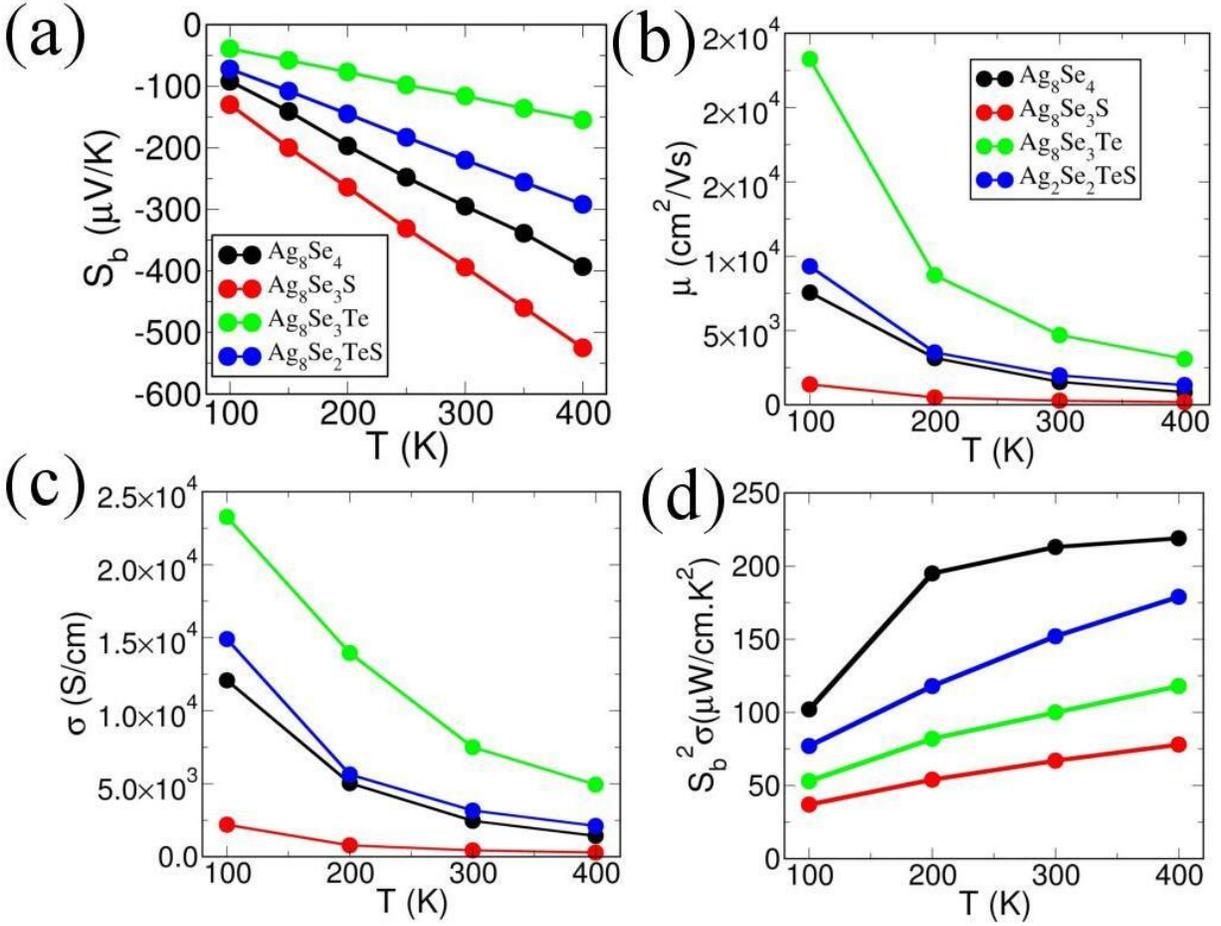

**Fig 5:** (a) Seebeck coefficient, (b) Mobility, (c) Electrical conductivity, (d) Power factor, have been shown in the temperature range (50-400 K). The Seebeck coefficient is derived from Boltzmann transport formalism. On the other hand, temperature-dependent mobility is calculated along $\Gamma \to K$ direction using acoustics phonon limited deformation potential theory. Finally, the Drude model gives the temperature-dependent electrical conductivity along $\Gamma \to K$ direction using the formula $\sigma = ne\mu_\beta^c$.

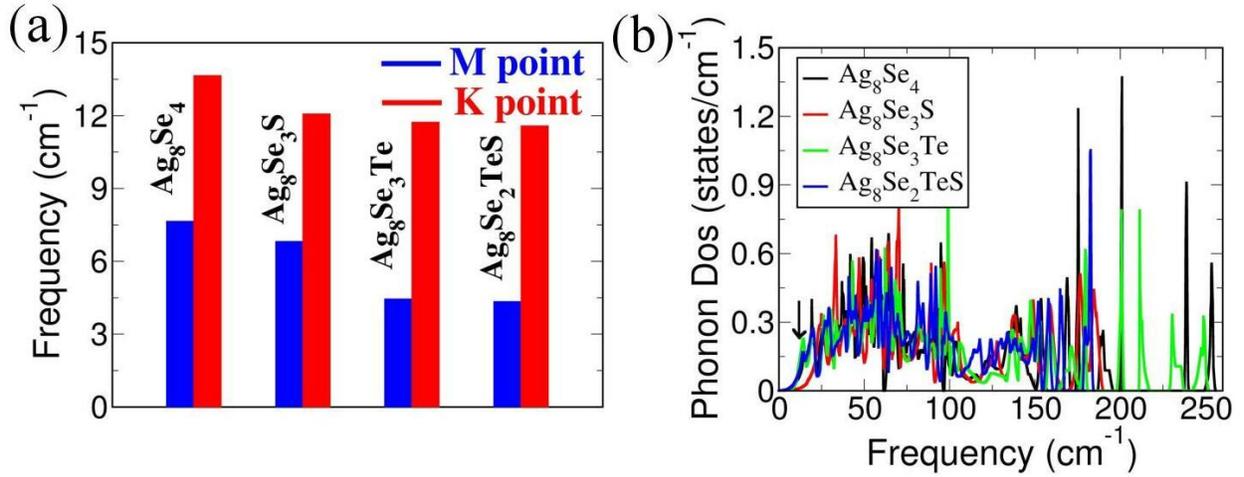

**Fig. 6:** (a) Soft phonon modes observed at both M-point (blue bar) and K-point (red bar) for pristine and doped $Ag_2Se$ in the Brillouin zone. Phonon frequency gradually decreases for $Ag_8Se_4$, $Ag_8Se_3S$, $Ag_8Se_3Te$, $Ag_8Se_2TeS$, respectively. (b) Phonon Dos for pristine and doped Ag2Se. The sharp peak indicated by the black ellipse signifies soft phonon modes at M and K points in the Brillouin zone for $Ag_8Se_3Te$ (green line) and $Ag_8Se_2TeS$ (blue line) systems.

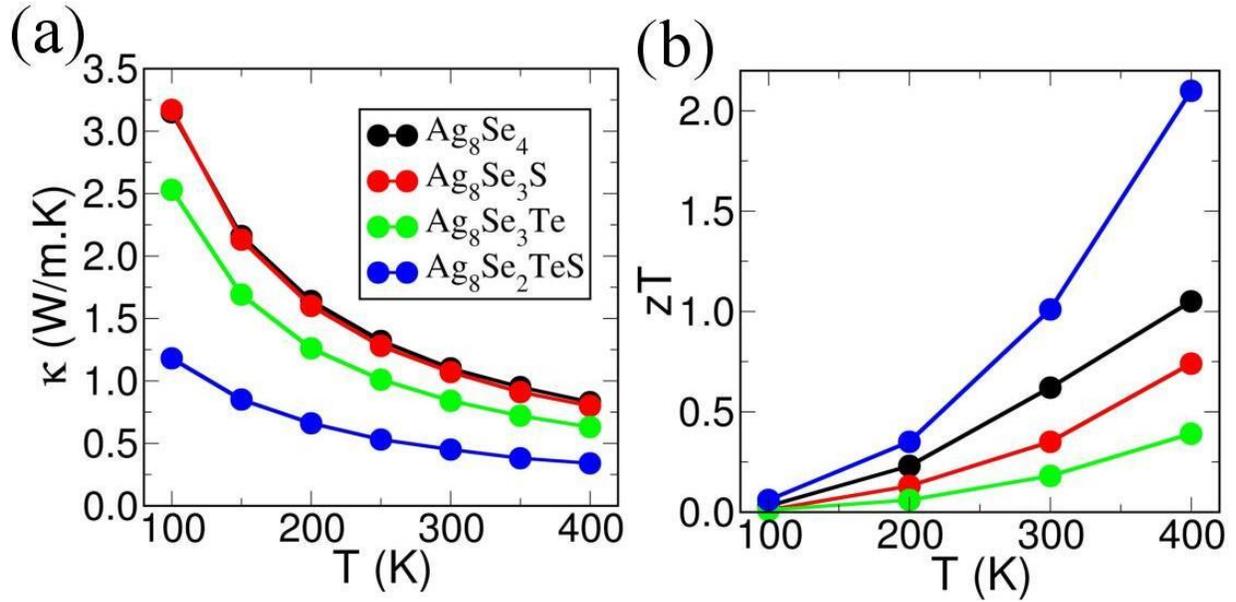

**Fig. 7:** Temperature-dependent (a) Total lattice thermal conductivity ($k$), (b) Figure of merit (zT), have been shown for pristine and doped $Ag_2Se$. The pseudoternary phase ($Ag_8Se_2TeS$) shows very high zT~2.6 at 400 K.

**Table 1:** Momentum offset (Å$^{-1}$) in the Brillouin zone. Pristine Ag$_2$Se behaves as a direct bandgap semiconductor whose VBM and CBM coincides at Γ-point (considered as reference) in the electronic dispersion curve.

| Displeasement (Å$^{-1}$) | Ag$_8$Se$_4$ | Ag$_8$Se$_3$S | Ag$_8$Se$_3$Te | Ag$_8$Se$_2$TeS |
|---|---|---|---|---|
| VBM | 0.0 | 0.005 | 0.0253 | 0.027 |
| CBM | 0.0 | 0.001 | 0.0253 | 0.037 |

**Table 2:** Change in configurational entropy ($\Delta S_{conf}$) at different doping concentrations of S and Te atoms in Ag$_2$Se unit cells.

| Change in configurational entropy | Ag$_8$Se$_4$ | Ag$_8$Se$_3$S | Ag$_8$Se$_3$Te | Ag$_8$Se$_2$TeS |
|---|---|---|---|---|
| $\Delta S_{conf}$ (Joule/K-mol) | 4.827 | 5.276 | 6.658 | 6.946 |

# Supporting Information

# Realizing high Near-Room-Temperature Thermoelectric Performance in n-type Ag$_2$Se through Rashba Effect and Entropy Engineering

Raju K Biswas and Swapan K Pati*

*Theoretical Sciences Unit*
*School of Advanced Materials*
*Jawaharlal Nehru Centre for Advanced Scientific Research (JNCASR)*
*Jakkur P.O., Bangalore 560064, India. *E-mail: pati@jncasr.ac.in*

===============================================================

Formation energy: $F.E = E(M@Ag_2Se) + E(Se) - E(Ag_2Se) - E(M)$

Binding energy: $B.E = E(Ag_2Se_{1-x}M_x) - E(Ag_2Se_{1-x}) - E(M)$

Here, M= Te and S atoms. $E(M@Ag_2Se), E(Se), E(Ag_2Se), E(M)$, are the total energies of unit cell where M atom substituted one Se atom, single Se atom in the unit cell, $Ag_2Se$ unit cell, and single M atom in the unit cell, respectively. On the other hand, $E(Ag_2Se_{1-x}M_x), E(Ag_2Se_{1-x}), E(M)$, are total energy of unit cell where one M atom substituted Se atom, one Se atom absent $Ag_2Se$ unit cell, and single M atom unit cell, respectively.

**Table S1:** Tabulated formation energy and binding energy for pristine and doped Ag2Se. Formation energy and binding energies for pristine Ag2Se are considered as reference. The negative values of formation energies and binding energy imply doped Ag2Se crystal structures are stable.

| Materials | $Ag_8Se_4$ | $Ag_8Se_3S$ | $Ag_8Se_3Te$ | $Ag_8Se_2TeS$ |
|---|---|---|---|---|
| Formation Energy (F.E) | 0.0 eV | -0.81 eV | -0.27 eV | -0.40 eV |
| Binding Energy (B.E) | 0.0 eV | -6.12 eV | -5.02 eV | -10.47 eV |

**Table S2:** List of k-points to plot electronic band structure for pristine and doped $Ag_2Se$:

| | | | |
|---|---|---|---|
| Γ | 0.0 | 0.0 | 0.0 |
| M | 0.5 | 0.0 | 0.0 |
| K | 0.5 | 0.5 | 0.0 |
| Γ | 0.0 | 0.0 | 0.0 |
| Z | 0.0 | 0.0 | 0.5 |
| R | 0.5 | 0.0 | 0.5 |
| X | 0.5 | 0.5 | 0.5 |
| Z | 0.0 | 0.0 | 0.5 |

**Table S3:** HSE06 estimated band gap for pristine and doped $Ag_2Se$.

| Materials | $Ag_8Se_4$ | $Ag_8Se_3S$ | $Ag_8Se_3Te$ | $Ag_8Se_2TeS$ |
|---|---|---|---|---|
| $E_g(eV)$ | 1.07 | 1.22 | 0.55 | 1.12 |

**Table S4:** $E^c$ and $C_{3D}$ obtained for hole under acoustics phonon approximation.

| Materials | $Ag_8Se_4$ | $Ag_8Se_3S$ | $Ag_8Se_3Te$ | $Ag_8Se_2TeS$ |
|---|---|---|---|---|
| $E^c(eV)$ | 17.2 | 1.68 | 2.09 | 1.32 |
| $C_{3D}(J/m^3)$ | $119 \times 10^{11}$ | $2.74 \times 10^{11}$ | $3.57 \times 10^{11}$ | $1.47 \times 10^{11}$ |
| $m^*$ | 0.78 | 0.96 | 0.81 | 0.51 |

**Table S5:** $E^c$ and $C_{3D}$ obtained for electron under acoustics phonon approximation.

| Materials | $Ag_8Se_4$ | $Ag_8Se_3S$ | $Ag_8Se_3Te$ | $Ag_8Se_2TeS$ |
|---|---|---|---|---|
| $E^c(eV)$ | 38.57 | 2.24 | 1.14 | 2.4 |
| $C_{3D}(J/m^3)$ | $119 \times 10^{11}$ | $2.74 \times 10^{11}$ | $3.57 \times 10^{11}$ | $1.47 \times 10^{11}$ |
| $m^*$ | 0.28 | 0.36 | 0.21 | 0.14 |

**Table S6:** $E^c$ and $C_{3D}$ obtained for phonon under acoustics phonon approximation.

| Materials | $Ag_8Se_4$ | $Ag_8Se_3S$ | $Ag_8Se_3Te$ | $Ag_8Se_2TeS$ |
|---|---|---|---|---|
| $E^c$ (eV) | 0.017 | 0.002 | 0.0029 | 0.0033 |
| $C_{3D}$ (J/m³) | 119x10¹¹ | 2.74x10¹¹ | 3.57x10¹¹ | 1.47x10¹¹ |

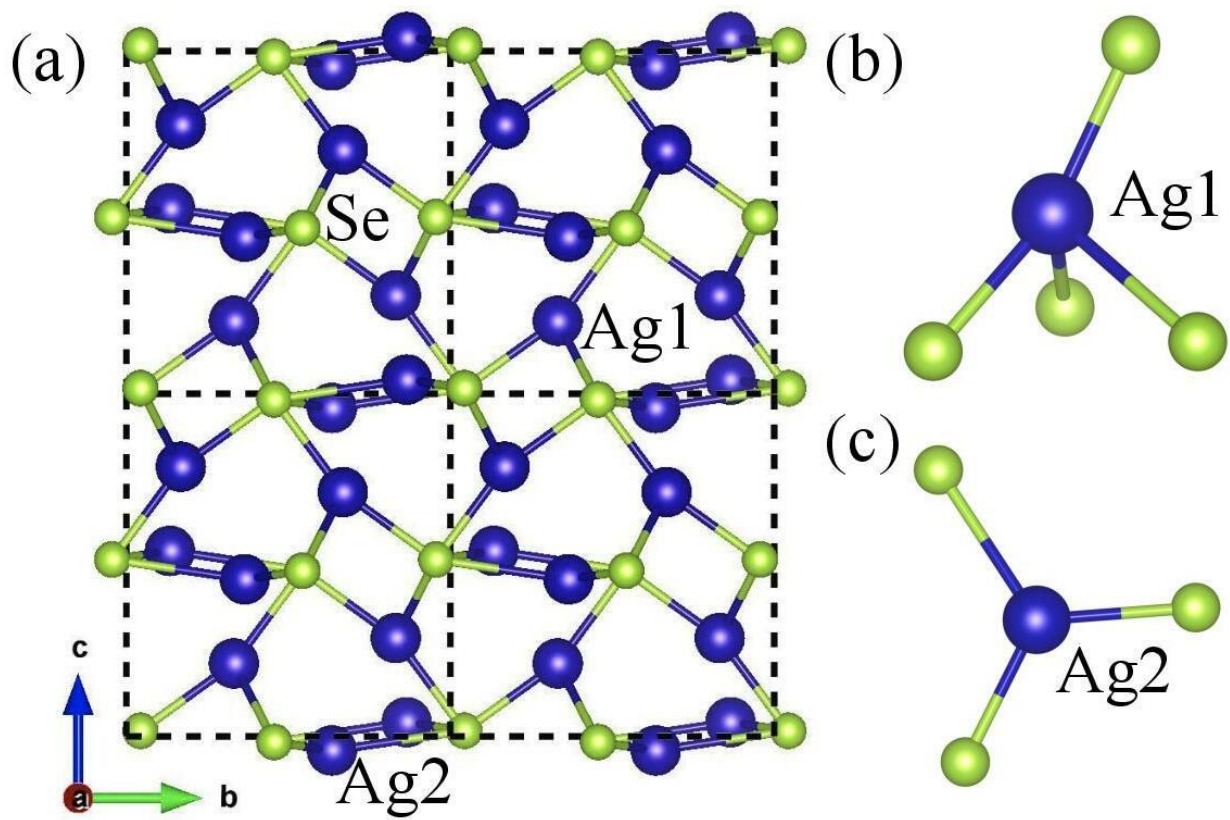

**Figure S1.** Crystal structure of $Ag_2Se$ at room temperature. Silver atoms are shown as blue cycles, selenium atoms as larger light green cycles. (a) $Ag_2Se$ crystal structure viewed from a-axis. Local atomic coordination of Ag atoms in the crystal structure where (b) Ag1 atoms are arranged tetrahedrally. (c) the coordination sphere of Ag2 is triangular.

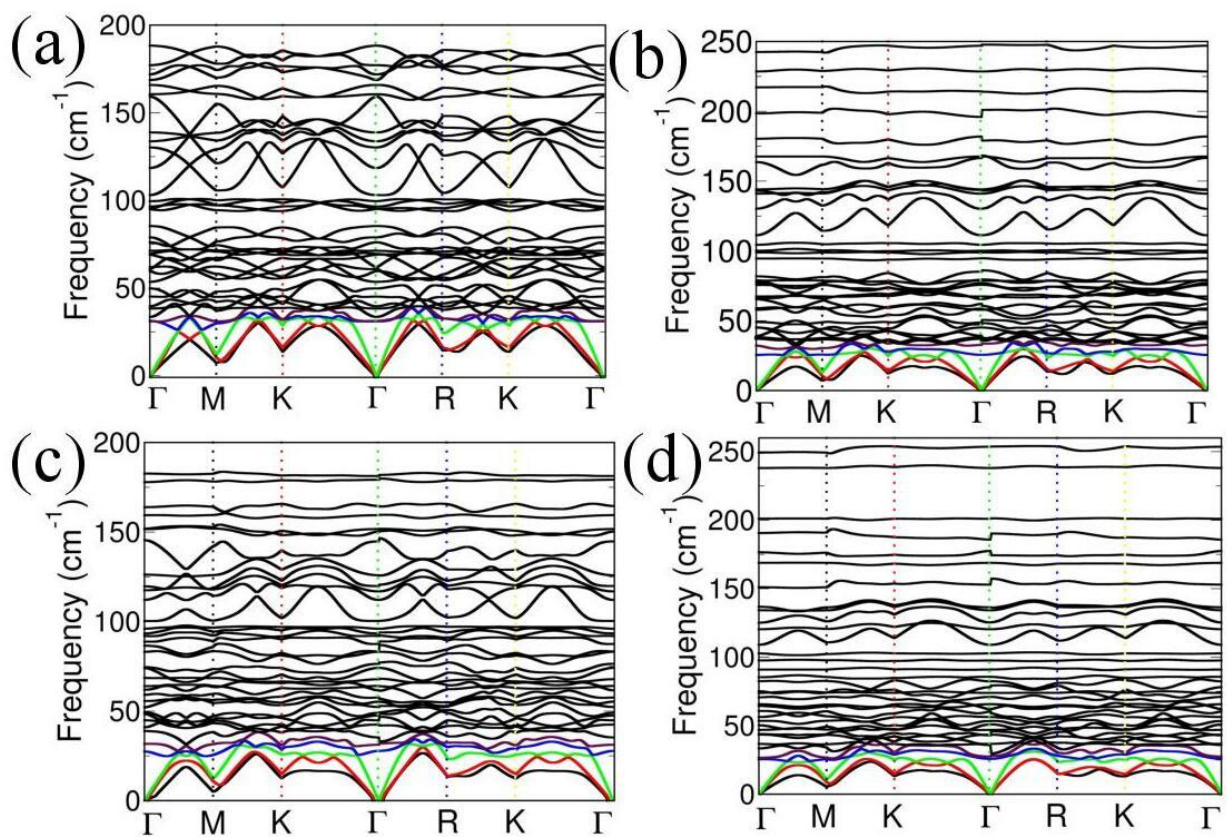

**Figure S2.** Phonon dispersion curve of (a) $Ag_8Se_4$, (b) $Ag_8Se_3S$, (c) $Ag_8Se_3Te$, (d) $Ag_8Se_2TeS$, have been calculated using DFPT and plotted as a function of q-point. The lattice dynamics curves signify both pristine and doped $Ag_2Se$ are dynamically stable. The green and red color phonon band represent longitudinal and transverse acoustics mode, respectively. On the other hand, blue color phonon band signifies low energetic optical phonon mode.

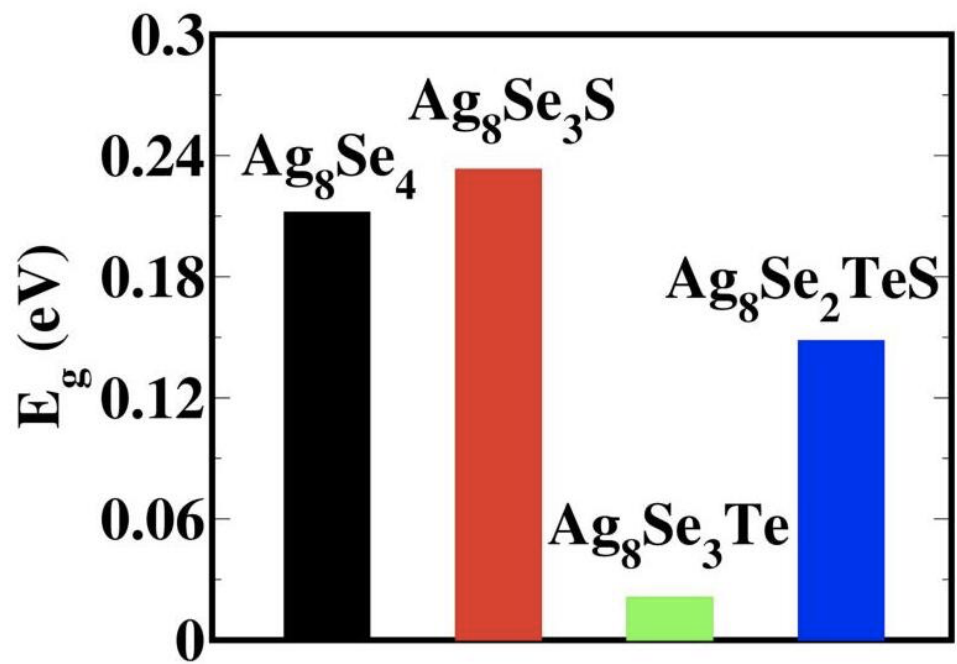

**Figure S3** PBE-SOC band gap for pristine and doped $Ag_2Se$.